\documentstyle[12pt,epsf,epsfig,a4wide]{article}
\begin{document}

\def\beq{\begin{equation}}
\def\eeq{\end{equation}}
\newcommand{\form}[1]{(\ref{#1})}

\begin{centering}
\begin{flushright}
CERN-TH/2002-033 \\
February 2002
\end{flushright}

\vspace{0.1in}

{\Large {\bf Strong Primordial Inhomogeneities and \\ Galaxy Formation}}

\vspace{0.4in}

{\bf M.Yu.~Khlopov$^{a,b}$, S.G.~Rubin$^{a,b}$} and {\bf A.S.~Sakharov$^{c,d}$
}
\\
\vspace{0.2in}

\vspace{0.4in}
 {\bf Abstract}

\end{centering}

\vspace{0.2in}

{\small \noindent The new element of theory of galaxy formation, strong
primordial inhomogeneities, is shown to be a reflection of
unstable large scale structures of topological defects, created in
second order phase transitions in the inflationary Universe.  In
addition to {\it archioles-like} large scale correlation of
the primordial inhomogeneity of energy density of coherent scalar field
oscillations, the same mechanism, based on the second order phase
transitions on the inflational stage and the domain wall formation upon
the end of inflation, leads to the formation of massive  black hole
clusters that can serve as nuclei for the future  galaxies. The
number of black holes with $M \sim 100M_{\odot}$  and above is
comparable with the number of galaxies within the modern
cosmological horizon. The primordial fractal structure of galaxies can
find natural grounds in the framework of model we developed  . The
proposed approach offers the physical basis for new scenarios of
galaxy formation in the Big Bang Universe.}

\vspace{0.8in}
\begin{flushleft}
CERN-TH/2002-033 \\
February 2002
\end{flushleft}

\vspace{0.5in}
\begin{flushleft} 

$^a$ Center for CosmoParticle Physics "Cosmion",  4 Miusskaya pl., 125047 
Moscow, Russia
\\
$^b$ Moscow Engineering Physics Institute, Kashirskoe shosse 31, 115409
Moscow, Russia \\
$^c$ Theory Division, CERN, CH--1211 Geneva 23 \\
$^d$ Swiss Institute of Technology,
ETH-H\"onggerberg, HPK--Geb\"aude, CH--8093 Zurich

\end{flushleft}


\newpage

\section{INTRODUCTION}

The origin of galaxies and of the observed structure of their
inhomogeneous distribution in FRW cosmology is generally ascribed
to the development of gravitational instability on the matter - dominant
stage, leading to the slow growth of small initial density fluctuations
on the nearly homogeneous expanding background. According to modern 
cosmology the spectrum of initial density fluctuations is
generated  on the inflationary stage. Particle theory offers 
a physical basis for  both inflation and dark matter, dominating in
the development of gravitational instability. However, the impact of
particle physics in the problem of galaxy formation seems to be much
stronger, adding new non-trivial elements to the theory of
cosmological structure formation. One such new element, which is a rather general 
cosmological consequence of the particle symmetry
breaking pattern, is the structure of strong primordial large scale
inhomogeneities, tracing the large scale structure of cosmological
defects, generated in second order phase transitions in the
inflationary Universe.

The existence of topological defects reflects the non-trivial
topological properties of broken particle symmetry. So, a magnetic
monopole solution follows from compact Lie group breaking, leaving
unbroken the U(1) symmetry of electromagnetism, which is the general
case for GUT models. In some of these, the conditions for cosmic
string and domain wall solutions are realized. The cosmological
effect of these topologically stable solutions is widely recognized
as an effective tool to probe the scenarios of very early Universe
and the particle models underlying them. Recall that magnetic monopole
overproduction~\cite{kz} has put severe constraints
on GUT models and stimulated the creation of inflationary
cosmology~\cite{guth}, whereas the domain wall problem \cite{okun} constrained 
the models of spontaneous CP
violation \cite{lee}. On the other hand, topological defects
represent a new element of galaxy formation, as it takes place in
cosmic string cosmology~\cite{zv}.

However, the actual amount of primordial magnetic monopoles produced in the 
inflationary Universe is still a big question, whereas
the conditions for stable wall and string solutions are not very
general for particle models (see the  
review \cite{khlop99}). A much
wider class of these models possesses the symmetry breaking
pattern, which can be effectively described by
pseudo-Nambu--Goldstone (PNG) field and which corresponds to the
formation of an unstable topological defect structure in the early
Universe.

The Nambu--Goldstone nature in such an effective description reflects
the spontaneous breaking of global symmetry, resulting in
continuous degeneracy of vacua. The explicit symmetry breaking at
smaller energy scale changes this continuous degeneracy by discrete
vacuum degeneracy. At high temperatures such a symmetry breaking
pattern implies the succession of second order phase transitions.
In the first transition, continuous degeneracy of vacua leads, at
scales exceeding the correlation length, to the formation of
topological defects in the form of a string network; in the
second phase transition, continuous transitions in space between
degenerated vacua form the surfaces: domain walls surrounded by
strings. This last structure is unstable, but,
as was shown in the example of the invisible axion \cite{kss}, it is
reflected in the large scale inhomogeneity of distribution of energy
density of coherent PNG (axion) field oscillations. The role of
inflation in the creation of this inhomogeneity is indirect, since it
provides identical initial conditions of expansion in causally
disconnected regions, which provides the simultaneous high
temperature phase transition and the formation of structure of
topological defects (axionic strings) that spreads far beyond the
cosmological horizon in that period.

Second order phase transitions on the inflational stage result in more
non-trivial influence of superluminal expansion on the forms of
topological defects and properties of their structure. The result
depends on the relationship between the symmetry breaking scales
and Hubble constant on inflational stage; it leads to various
observational consequences, related to various strong primordial
inhomogeneities. Among these consequences, systematic study of
which we approach, there appears a new interesting way of formation of primordial 
galactic nuclei, to be discussed in the present paper.

Now there is no doubt that the centres of almost all galaxies
contain massive black holes \cite{Rosen}. An original explanation
of the formation of such supermassive black holes assumes the
collapse of a large number of stars in the galaxy centres. However,
the mechanism of the galactic nuclei formation is still unclear.
According to \cite{Veil}, there are serious grounds to
believe that the formation of stars and galaxies proceeded
simultaneously.  
On the other hand, in the work \cite{Stia}  a model of galaxy formation around a 
massive black hole was
considered and arguments in its favour were presented. 
Each of the two approaches has
certain advantages, while being neither free of drawbacks.

We will consider below a new mechanism describing the very early
formation of primordial black holes (PBHs), which serve as the nucleation 
centres in the
subsequent formation of galaxies. This mechanism may prove to be
free from disadvantages inherent in the models based on the concept
of a single PBH being the nucleus of the future galaxy.

Previously \cite{Ru2} we proposed a mechanism of the PBHs
formation that opens the possibility of massive black hole
formation in the early Universe. The mechanism is based on the
fact that black holes can be created as a result of a collapse
of closed vacuum walls formed during a second order phase transition. The
masses of such black holes may vary within broad limits, up to the
level $\sim 10^8M_{\odot}$.

To figure out the idea \cite{Ru2}, let us assume that a potential
of some scalar field possesses at least two different vacuum
states. In such a situation there are two quantitatively different
possibilities to distribute these states in the early Universe. The
first possibility implies that the Universe contains approximately
equal amounts of both vacuum states, which is typically the case at
the usual thermal phase transition. The alternative possibility
corresponds to the case when the two vacuum states are populated
with different probabilities. In this case, islands of a less
probable vacuum state surrounded by a sea of another, more
probable vacuum state appear. As was shown in \cite{Ru2}, an
important condition for such an asymmetric distribution is the
existence of effectively flat valleys in the scalar field potential
during inflation. Under this condition the scalar field can be
considered as a massless scalar field  additional to inflaton,
existing on the inflational de Sitter background. It is well known
that the inflational fluctuations can essentially change the value
of a massless scalar field, while the
inflation itself blows up wavelengths of these fluctuations. These
two factors define the space distribution of a scalar field, which
is still massless until a certain moment. Thus, even though the
phase transition itself takes place only after the end of inflation,
deeply in the Friedman--Robertson--Walker (FRW) epoch, the initial
distribution of the scalar field is already defined at the
inflation period, 
in such a way that there are eventually, islands
representing  one vacuum in the sea of another vacuum.

It is supposed that after the phase transition, the two vacuum
states are separated by a vacuum wall. The initial distribution of
the scalar field formed during the inflation stage allows the formation of
closed walls of sizes significantly exceeding the cosmological
horizon at the moment when the phase transition just takes place.
At some instant after crossing the horizon, such walls become
causally connected as a whole and begin to contract because of the
surface tension. As a result, provided that friction is small and
the wall does not radiate a considerable part of its energy in the
form of scalar waves, almost all the energy of a closed wall may be
focused within a small volume inside the gravitational radius. This
is the necessary condition for a black hole formation. The mass
spectrum of black holes formed by this mechanism depends on
parameters of the scalar field potential determining the direction
and size of the potential valley during inflation and the
post-inflation phase transition. Although we deal here with the so-called  PNG 
field, the proposed mechanism
is quite general. The presence of massive PBHs is a new factor in
the development of gravitational instability in the surrounding
matter and may serve as a basis for new scenarios of the formation and
evolution of galaxies.

 \section {PRIMORDIAL BLACK HOLE FORMATION}
Now we will describe a mechanism accounting for the appearance of
massive walls of a size essentially greater than the horizon at
the end of inflation. Let us consider a complex scalar field with
the potential
\begin{equation}\label{V1} V(\varphi ) = \lambda (\left| \varphi
\right|^2  - f^2 /2)^2+\delta V(\theta ), \end{equation}
where $\varphi  = re^{i\theta } $. This field coexists with an
inflaton field which drives the Hubble constant $H$ during
the inflational stage. The term
\begin{equation} \label{L1} \delta V(\theta ) = \Lambda ^4 \left(
{1 - \cos \theta } \right), \end{equation}
reflecting the contribution of effects to the Lagrangian
renormalization (see for example \cite{adams}), is negligible on the 
inflational stage and during some period in the FRW expansion. In
 other words, the parameter  $\Lambda$  vanishes with respect to
$H$. The omitted term (\ref{L1}) begins  to play a
significant role only at the moment, after inflation, when the Hubble
parameter sharply decreases with time ($H = 1/2t$ during the
radiation dominated epoch). Also, we assume the mass of the radial
field component $r$ always to be sufficiently large with respect to
$H$, which means that the complex field is in the ground
state even before the end of inflation. Since the term (\ref{L1})
is negligible during inflation, the field has the form $\varphi
\approx f/\sqrt 2 \cdot e^{i\theta } $, the quantity $f\theta$
acquiring the meaning of a massless field.

At the same time,  the well established behaviour of quantum field
fluctuations on the de Sitter background \cite{Star79} implies that
the wavelength of a vacuum fluctuation of every scalar field grows
exponentially, having a fixed amplitude. Namely, when the
wavelength of a particular fluctuation, in the inflating Universe,
becomes greater than $H^{-1}$, the average amplitude of this
fluctuation freezes out at some  non-zero value because of the large
friction term in the equation of motion  of the scalar field,
whereas its wavelength grows exponentially. Such a frozen
fluctuation is equivalent to the appearance of a classical field that
does not vanish after averaging over macroscopic space intervals.
Because the  vacuum must contain fluctuations of every wavelength,
inflation leads to the  creation of more and more new regions
containing a classical field of different amplitudes with scale
greater than $H^{-1}$. In the case of an effectively massless
Nambu--Goldstone field considered here, the averaged amplitude of
phase fluctuations  generated during each e-fold (time interval
$H^{-1}$)  is given by
\beq \label{fluctphase} \delta \theta = H/2\pi f. \eeq
Let us assume that the part of the Universe
observed inside the contemporary horizon $H_0^{-1}=3000h^{-1}$Mpc
was inflating, over $N_U \simeq 60$ e-folds, out of a
single causally connected domain of size $H^{-1}$, which contains some
average value of phase $\theta_0$ over it. 
When inflation begins in this region, after one e-fold, the volume of the Universe 
increases by a factor $e^3$ . The typical wavelength of the fluctuation 
$\delta\theta$
generated during every e-fold is equal to $H^{-1}$. Thus, the whole
domain  $H^{-1}$, containing $\theta_{0}$, after the first e-fold
effectively becomes divided into  $e^3$  separate, causally
disconnected domains of size $H^{-1}$. Each domain contains almost
homogeneous  phase value $\theta_{0}\pm\delta\theta$. Thereby, more
and more domains appear with time, in which the phase differs
significantly from the initial value $\theta_0$. A principally
important point is the appearance of domains with phase
$\theta >\pi$. Appearing only after a certain period of time during
which the Universe exhibited exponential expansion, these domains
turn out to be surrounded by a space with phase $\theta <\pi$.
The coexistence of domains with phases $\theta <\pi$ and $\theta
>\pi$ leads, in the following, to the formation of
a large-scale structure of topological defects. 

The potential (\ref{V1}) possesses a $U(1)$ symmetry, which is
spontaneously broken, at least, after some period of inflation. Note that the 
phase fluctuations during the
first e-folds may, generally speaking, transform eventually into
fluctuations of the cosmic microwave radiation, which will lead to
imposing restrictions on the scaling parameter $f$. This difficulty
can be avoided by taking into account the interaction of the field
$\varphi$ with the inflaton field (i.e. by making parameter $f$ a
variable~\cite{progress}). This spontaneous breakdown is holding by the
condition of smallness of the radial mass,
$m_r=\sqrt{\lambda_{\phi}}>H$. At the same time the
condition
\beq\label{angularmass} m_{\theta}=\frac{2f}{\Lambda}^2\ll H \eeq
on the angular mass provides the freezing out  of the phase
distribution until some moment of the FRW epoch.  After the
violation of condition (\ref{angularmass}) the term (\ref{L1})
contributes significantly to the potential (\ref{V1}) and
explicitly breaks the continuous symmetry along the angular
direction. Thus, potential (\ref{V1}) eventually has a number of
discrete degenerate minima in the angular direction at the points
$\theta_{min}=0,\ \pm 2\pi ,\ \pm 4\pi,\ ...$ .

As soon as the angular mass $m_{\theta}$ is of the order of the
Hubble rate, the phase starts oscillating about the potential
minimum, initial values being different in various space
domains. 
Moreover, in the domains with the initial phase $\pi <\theta <
2\pi $, the oscillations proceed around the potential minimum at $\theta
_{min}=2\pi$, whereas the phase in the surrounding space tends to a
minimum at the point $\theta _{min}=0$. Upon ceasing of the
decaying phase oscillations, the system contains domains
characterized by the phase $\theta _{min}=2\pi$ surrounded by space with $\theta 
_{min}=0$. Apparently, on moving in any
direction from inside to outside of the domain, we will
unavoidably pass through a point where $\theta =\pi$ because the phase varies 
continuously. This implies that a
closed surface characterized by the phase $\theta _{wall}=\pi$ must
exist. The size of this surface depends on the moment of domain
formation in the inflation period, while the shape of the surface
may be arbitrary.  
The principal point for the subsequent
considerations is that the surface is closed.
After reheating of the Universe, the evolution of domains with
the phase $\theta >\pi $ 
proceeds on the background of the Friedman
expansion and is described by the relativistic equation of state.
When the temperature falls down to $T_* \sim \Lambda$, 
an equilibrium state between the "vacuum"
phase $\theta_{vac}=2\pi$ inside the domain and the $\theta_{vac} =0$
phase outside it is established. Since the
equation of motion corresponding to potential \form{L1} admits a
kink-like solution (see \cite{vs} and references therein), which
interpolates between two adjacent vacua $\theta_{vac} =0$  and
$\theta_{vac} =2\pi$,  a closed wall corresponding to the
transition region at $\theta =\pi$ is formed.   
The surface energy
density of a wall of width $\sim 1/m\sim  f/\Lambda^2$ is
of the order of $\sim f\Lambda ^2$ \footnote{The existence of such domain walls 
in theory of the invisible axion was first pointed out in \cite{sikivieinvisible}.}.

Note that if the coherent phase oscillations do not decay for a
long time, their energy density can play the role of CDM. This is
the case, for example, in the cosmology of the invisible axion
(see \cite{kim} and references therein).

It is clear that immediately after the end of inflation, the size of
domains which contains a phase $\theta_{vac} >2\pi$ essentially
exceeds the horizon size.  This situation is replicated in the size
distribution of vacuum walls, which appear at the temperature $T_*
\sim \Lambda$ whence the angular mass $m_{\theta}$ starts to build
up. Those walls, which are larger than the cosmological horizon, still
follow the general FRW expansion until the moment when they get
causally connected as a whole; this happens as soon as the size of
a wall becomes equal to the horizon size $R_h$. Evidently,
internal stresses developed in the wall after crossing  the horizon
initiate processes tending to minimize the wall  surface. This
implies that the wall tends, first, to acquire a  spherical shape
and, second, to contract toward the centre. For simplicity,
we will consider below the motion of closed spherical walls~\footnote{The
motion of closed vacuum walls has been driven analytically in
\cite{tkachev,sikivie}.}.

The wall energy is proportional to its area at the instant of
crossing the horizon. At the moment of maximum contraction, this
energy is almost completely converted into kinetic energy.
Should the wall at the same moment be localized within the
gravitational radius, a PBH is formed.

Detailed consideration of BH formation was performed in~\cite{Ru28}. We proceed 
below to study the formation of a PBH
cluster in the
early Universe.

\section{CORRELATIONS IN THE BLACK HOLE \\ DISTRIBUTION}

Previously  \cite{Ru2,Ru28} we had studied a new process involving the
formation of uncorrelated PBHs in the Universe. It was demonstrated
that a model with reasonable parameters readily provides for the
formation of $\sim 10^{11}$ massive (with the mass $10^{30}$-$10^{40}$g. each)
black holes, which coincides with  the number of galaxies in the
observed part of the Universe.
In that analysis, we did not take into account
correlations (inherent in this mechanism) between the formation of
a massive black hole and the appearance of smaller
black holes surrounding it.
This correlation is related primarily
to certain features of the above-discussed process of the formation
of domains with phases $\theta >\pi$. It seems that the
appearance of such domains creates prerequisites for the formation
of new, smaller domains inside. Let us estimate the mass
distribution of these daughter domains. Consider a region with a
size
of the order of $H^{-1}$ and a phase within $\pi<\theta_0<\pi+\delta$ (where
$\delta =H/2\pi f$  is the average phase jump
during the $H^{-1}$ time period), formed during the inflation period
as a result of fluctuation in a certain region of space of phase $\theta < \pi$. 
During the next e-fold, this space domain
will
be divided in $e^3$ subdomains with size $H^{-1}$, and
some of  these will acquire
a phase $\theta_1$ in the interval $\pi-\delta <\theta_1 <\pi$.
Upon the subsequent phase transition, these domains will be
separated by walls from the external region. Similar transitions,
when crossing the phase $\theta =\pi$ in the reverse direction, will
take place in each subdomain during the next e-fold. Thus, a
structure of the fractal type appears \footnote{A fractal structure of unclosed 
axionic walls, which is generated on the inflationary stage has been discussed in 
\cite{lindefractals}.} which reproduces itself in
each time step on a decreasing scale.

Let $\zeta$ denotes the number of subdomains formed in each step,
around which a wall may appear with time. Apparently, this value
obeys the inequality $1<\zeta\ll e^3$. In the subsequent estimates,
we will assume that $\zeta\approx 2$--$3$. Since each
causally connected domain touches approximately six neighboring
domains, we can hardly expect $\zeta$ to be greater for a total
number of $\sim e^3 \approx 20$. The mass of the future black hole
(if this would actually form) is determined by the area of a closed
surface with the phase $\theta =\pi$. The ratio of areas of the
initial (mother) and daughter domains is readily estimated: the
initial area, after a single e-fold, is $S_0 \approx e^2 H^{-2}$, and
the daughter subdomain area is $S_1 \approx H^{-2}$. Therefore, the
ratio of masses of the black holes belonging to two sequential
generations is
\begin{equation}
M_j /M_{j + 1}  \approx S_j /S_{j + 1}  \approx e^2,
\end{equation}
for a relative number of them assumed to be
\begin{equation}
N_{j + 1} /N_j  = \zeta .
\end{equation}
As is readily seen, the number and mass of black holes appearing
upon the $j$-th e-fold after the initial domain formation are
determined by parameters of the largest black hole genetically
related to the primary domain in which the phase originally
exceeded $\pi$. It is evident that
\begin{equation}
N_j  \approx \zeta ^j ;\quad M_j  \approx M_0 /e^{2j}.
\end{equation}
Excluding $j$ from these relationships, we obtain the desired black
hole mass distribution in a cluster:
\begin{equation}
N_{cl} (M) \approx \left( {M_0 /M} \right)^{\frac{1}{2}\ln \zeta }.
\end{equation}
The total mass of the cluster can be expressed through the mass
$M_0$ of the largest initial black hole:
\begin{equation}
\begin{array}{l}
 M_{tot}  = M_0  + \zeta M_1  + \zeta ^2 M_2  + ... =  \\
 M_0  + \zeta e^{ - 2} M_0  + \left( {\zeta e^{ - 2} } \right)^2 M_0  +
... = M_0 [1 - \zeta /e^2 ]^{-1}. \\
  \end{array}
\end{equation}
As is seen, the total mass of the black hole cluster is only one
and a half to two times greater than the largest initial black hole
mass. The number of daughter black holes depends on the factors
considered in the next section.

The inflationary mechanism described above leads to the occurrence
of a fractal structure of the closed walls. After the end of
inflation, as soon as the size of the horizon becomes larger than the
characteristic size of closed walls, the walls begin to shrink. The
energy of each wall, proportional to the area of their surface,
concentrates in small spatial domains (in the following they are
considered as point-like objects)\cite{Ru2, Ru28}. These high density
clots of energy could serve in the following for star and/or galaxy
formation (see for example \cite {khlop99} and references there). Hence, 
according to the given models,
the distribution of stars and galaxies should have fractal
properties as well. It is important to note that the total surface
of the walls in definite volume is proportional to the total energy of
an object, while the number of walls is equal to the number of
dense clots.

Suppose, for a characteristic time $1/H$, that several closed walls appear
in a causally connected area of size $R$, and the phase
is still at the maximum 
of the
potential. Denote the number of walls by $N$ and their average size
by $\xi R$, $\xi > 1/e$ ($ \xi \neq 1/e$ because of a possible merging
of causally disconnected subdomains with one common wall). In each
of these subdomains, $N$ new, smaller closed walls of size $\xi^{2}
R$ arise during the next time step. Denote by $a$ the minimal
size of such a wall that we are able to distinguish. This means
that we may terminate the process after the $n$-th step such that $a
\equiv \xi^n R$. The total area of the closed walls in the initial
volume is the sum of the areas with closed walls of size greater than
$a$. This simple summation leads to the following result:
\begin{equation}
\label {Frac1} S\approx R ^ 2 q (q ^ n -1) / (q-1),\quad q\equiv
\xi ^ 2 N ~ .
\end{equation}
This expression can be written in the form
\begin {equation}
\label {frac2} S\approx (R/a) ^ D ~ ,
\end {equation}
where $D$ is the fractal dimension. Equating these two expressions,
one obtains 
\begin {equation}
\label {Fracdim} D = 2 + \frac {{\ln \left ({q\frac {{q ^ {\frac
{{\ln \left ({a/R} \right)}} {{\ln \xi}}} - 1}} {{q - 1}}}
\right)}} {{\ln (R/a)}} ~ .
\end {equation}
This quantity is constant only when the ratio $R/a$ is large; it is
different for $ q < 1 $ and $ q> 1 $. It can be easily verified
that $D\rightarrow 2$ for $ q < 1$, while for $ q> 1 $,
$D\rightarrow 2 + 3\ln (q) /\ln (4N)$. To get an estimate, suppose
that the number of closed domains is $ N\approx 4$, and $\xi
\approx 1/e$. The value of the parameter $q$ can be easily
calculated, $q\approx 0.5$. Hence, the fractal dimension of the
system of closed walls $D\approx 2$.

So, if quantum fluctuations lead to the formation of spatial areas,
with the phase taking a value near a potential maximum, its further
evolution results in a system of enclosing walls. The
characteristic size of the next generations of walls differs from
the previous one by a factor of approximately $e$. The fractal
dimension of such a system is $D\approx 2$.

According to this scenario, it is interesting to find the number of
walls inside a sphere of radius $R$, which is given by
\begin {equation}
\label {Ntot} N _ {tot} = \sum _ {i = 1} ^ {n} N ^ {i} = N\frac {N
^ n -1} {N-1} \approx \frac {N ^ {n + 1}} {N-1} ~ .
\end {equation}
By analogy with the previous calculations and using
Eq. (\ref{Ntot}), one obtains the distribution  of point--like dense
objects with fractal dimension $D' \approx \ln N/\ln (1/\xi)$. For the realistic 
values $N\approx 4 and \quad \xi \approx 1/e$, we find $D'
\approx 1.4$, which differs somewhat from the value $D \approx 2$
previously obtained. This is not surprising since in the first
case we measure the area of wall surfaces within a certain
volume, while in the second case we measure the number of walls.

Let us compare our calculations with observational data of spatial
distribution of galaxies and of stars in those galaxies. Recent
data indicate that the distribution of stars and galaxies indeed
carries a fractal character. So the number of galaxies inside a
sphere of radius $R$ is $N(R)\sim (R)^{2.2\pm 0.2}$ up to a sizes
of 200 Mpc \cite {Labini98}.

The distribution of stars inside galaxies also carries a fractal
character. In Ref.\cite {FractStar} this fractal dimension was
determined by averaging observational data from ten galaxies; it was
found to be equal to $D \sim 2.3$.

Evidently, the observable fractal dimension $D$ in distributions of
stars and galaxies is in agreement with predictions of our model. Of course, 
other mechanisms, at a later stage, may contribute
to the distribution and change the fractal dimension somewhat, but the scenario  
of PBHs formation, discussed here, could give a primordial reason of fractality 
in the galaxy and star distribution.

The mechanism of fractal structure production discussed in this
paper is not unique. Another example is based on hybrid inflation,
one of the most promising models of inflation \cite{Linde91a,
Dvali94, Luth99}. In the standard version of hybrid inflation the
potential contains two fields 
\begin{equation}
V = V_0  + \frac{1}{2}m_\varphi ^2 \varphi ^2  + \frac{1}{2}\lambda
_1 \varphi ^2 \psi ^2  - \frac{1}{2}m_\psi ^2 \psi ^2  +
\frac{1}{2}\lambda _2 \psi ^4 ~ .
\end{equation}
During inflation, the field $\varphi$ rolls down along a valley
$\psi =0$. In the meantime field fluctuations around the critical
line $\psi =0$ lead to the formation of a fractal structure of domains.
This critical line plays the same role as the critical point $\pi$
in the previous discussion.  Just after passing the critical point
$\varphi =m_{\psi}^2/\lambda _1$, the state $\psi =0$ becomes
unstable and field $\psi$ moves (on average) to one of the new
stable minima. These minima are separated by a potential maximum and
we again inevitably come to the fractal structure of domain walls.
\begin{figure}
\begin{center}
\epsfig{file=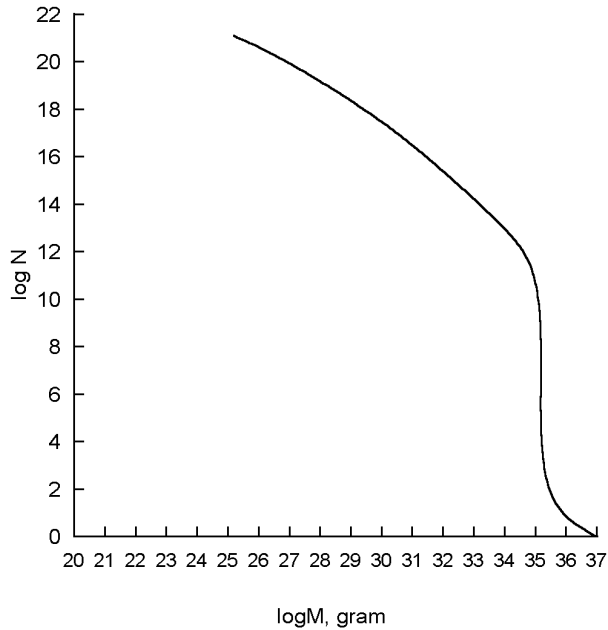,width=80mm, clip=}
\end{center}
\caption{\it The numerically simulated distribution of PBHs, where $f = 1.77H$  
and $\Lambda = 5$GeV. It is supposed that the Universe underwent 60 e-folds 
of inflational expansion with the Hubble constant  $H = 10^{13}$GeV.}
\label{fig:distribution}
\end{figure}

\section{DISCUSSION}

In the preceding sections, we considered only the principal
possibility of the formation of domain walls connecting adjacent
vacuum states. We have used the formulas derived above to estimate
the efficiency of the proposed mechanism of the black hole cluster
formation. 
The numerical calculations \cite{Ru28} were performed for the
following values of the parameters (which are consistent with the
observed anisotropy in the cosmic microwave radiation): the Hubble
constant at the end of inflation, $H = 10^{13}$Gev; Lagrangian
parameters, $f = 1.77H$  and $\Lambda = 5$ GeV. The initial phase,
at which the visible part of the Universe is formed between the time
$t_U \approx 60H^{-1}$ to the end of inflation, controls the number
of domains and, accordingly, the number of closed walls formed in
the post-inflation stage. This random value, not related to the
Lagrangian parameters, must be selected, taking into account the
numerous restrictions on the abundance of PBHs in the
Universe (see for example \cite{khlop99} and references there). We will use the 
numerical value $\theta _U =0.05\pi$,
which ensures a sufficiently large number of massive black holes,
while the presence of numerous smaller ones does not
contradict experimental restrictions.
 As is seen in Fig.\ref{fig:distribution}, the
PBH masses fall within the range from $10^{25}$ to $10^{35}$g. The
initial phase $\theta_U$ was selected so as to provide for the
number of massive PBHs (with mass $\sim 10^{35} g$) to be equal to the number
of galaxies in the visible part of the Universe. The total mass of
black holes  
is supposed to be $\sim 1\%$ of the
contemporary baryonic contribution.

The results of these calculations are sensitive to changes in the
parameter $\Lambda$ and the initial phase $\theta _U$. As the
$\Lambda$ value decreases to $\approx 1$GeV, still greater PBHs
appear with masses of up to $\sim 10^{40}$ g. A change in the initial
phase leads to sharp variations in the total number of black holes.
As was shown above, each domain generates a family of
subdomains in the close vicinity. The total mass of such a cluster is
only 1.5--2 times that of the largest initial black hole in this
space region. Thus, our calculations confirm the possibility of
formation of clusters of massive PBHs ( $\sim 100M_{\odot}$ and
above) in the earliest stages of the evolution of the Universe at a
temperature of $\sim 1-10$GeV. These clusters represent stable
energy density fluctuations around which increased baryonic 
(and cold dark matter) density
may concentrate in the subsequent stages, followed by the evolution
into galaxies.

It should be noted that additional energy density is supplied by
closed walls of small sizes. Indeed, because the smallness of their
gravitational radius, they do not collapse into BHs. After several
oscillations such walls disappear, leaving coherent fluctuations of
the PNG field. These fluctuations contribute to a local energy
density excess, thus facilitating the formation of galaxies.

\section{CONCLUSIONS}

This paper proposes a new mechanism for the formation of
protogalaxies, which is based on the cosmological inferences of the
elementary particle models predicting non-equilibrium second order
phase transition \cite{Ru2} in the inflation stage period and the 
formation of a domain wall upon the end of inflation. The presence of closed 
domain
walls of a size markedly exceeding the cosmological horizon 
in their period of formation leads to
the collapse of the wall in the post-inflation epoch (when the wall size becomes
comparable with the cosmological horizon); this results of a formation of
massive black hole clusters that can serve as nuclei for the future
galaxies. The mass spectrum of PBHs do
not contradict the available restrictions. The number of black
holes with $M \sim 100 M_{\odot}$ and above is comparable with
the number of galaxies in the visible Universe. Processes of deceleration of the 
wall motion, which have been considered in~\cite{Ru28}, may affect only the 
dynamics of collapse of
supermassive walls, which are beyond of the scope of this paper.

A development of the proposed approach gives ground for a principally
new scenario of the galaxy formation in the model of the hot Universe.
Traditionally, the hot Universe model assumes a homogeneous
distribution of matter on all scales, whereas the appearance of
observed inhomogeneities is related to the growth of small initial
density perturbations. 
However, the analysis of the cosmological
consequences of the particle theory indicates the
possible existence of strongly inhomogeneous primordial structures
in the distribution of both the dark matter and baryons. These
primordial structures represent a new factor in galaxy
formation theory. Topological defects such as the cosmological walls and
filaments, primordial black holes, archioles in the models of axionic
CDM \cite{kss}, and essentially inhomogeneous
baryosynthesis (leading to the formation of antimatter domains in
the baryon-asymmetric Universe)~\cite{barions} offer by no means a
complete list of possible primary inhomogeneities inferred from the
existing elementary particle models.

The proposed approach discloses a number of interesting aspects in
this direction. Indeed, our model offers a possibility of quantitative analysis 
of correlations in the formation of
massive PBHs and the primary inhomogeneity of dark matter and
baryons. Originally inherent in this mechanism is the inhomogeneous
phase distribution, which eventually acquires (much as what takes
place in the invisible axion cosmology) a dynamical sense of the
initial amplitude of the coherent oscillations of a scalar field.
Irrespective of the efficiency of dissipation of the energy of
these oscillations, the regions of closed wall formation must be
correlated with the regions of maximum energy density of the latent
mass. If these oscillations are not decaying, their energy density
may provide for the contemporary CDM density. Inhomogeneity
in the initial amplitude of these oscillations would then imply 
the inhomogeneity in the initial energy density and, hence, the regions
of black hole formation would become the regions of increased dark matter 
density.

A qualitatively similar effect (albeit not as pronounced) takes place
in the dissipation of coherent oscillations at the expense of
particle production. An increase in the oscillation energy density
transforms into a local increase in the density of latent mass
particles produced in this region.

The development of the proposed approach may thus lead to a number of
interesting scenarios of the initial stages in the formation of
protogalaxies, depending on the selection of particular elementary
particle models and their parameters. This study presents the first
step in this direction.

\section{ACKNOWLEDGMENTS}

The authors are grateful to E.D.~Zhizhin and A.A.~Starobinsky for
their interest in this study. The work of 
M.Yu.Kh. and S.G.R. was
partly performed within the framework of the "Cosmoparticle
Physics" Project of the State Scientific - Technological Program
"Astronomy. Fundamental Space Research", and it was supported by the
Cosmion-ETHZ, Epcos-AMS international cooperations and Scientific
School Support Grant no. 00-15-96699. One of us
(M.Yu.Kh.) expresses his gratitude to IHES and ICTP
for hospitality, while finishing the work on this paper. We are also grateful to 
J.~Ellis for useful comments.

\end{document}